\documentclass[twocolumn]{aa}
\usepackage{graphicx}
\begin{document}
\title{Large-scale galaxy correlations as a test for dark energy}

\titlerunning{Galaxy correlations test for dark energy}

\author{Alain Blanchard
\inst{1}
\and
Marian Douspis
\inst{2}
\and
Michael~Rowan-Robinson
\inst{3}
\and
Subir~Sarkar
\inst{4}
}
\offprints{Alain Blanchard}

\institute{
LATT, 14 avenue Edouard Belin, F-31400 Toulouse, France\\
\email{alain.blanchard@ast.obs-mip.fr}
\and
LATT, 14 avenue Edouard Belin, F-31400 Toulouse, France\\
\email{douspis@ast.obs-mip.fr}
\and
Astrophysics Group, Imperial College, Blackett Laboratory,
Prince Consort Road, London SW7 2BW, UK
\and
Rudolf Peierls Centre for Theoretical Physics, University of Oxford, 1
Keble Road, Oxford OX1 3NP, UK\\
\email{sarkar@thphys.ox.ac.uk}}

\date{}

\abstract{We have shown earlier that, contrary to popular belief,
Einstein--de Sitter (E--deS) models can still fit the {\sl WMAP} data
on the cosmic microwave background provided one adopts a low Hubble
constant and relaxes the usual assumption that the primordial density
perturbation is scale-free. The recent {\sl SDSS} measurement of the
large-scale correlation function of luminous red galaxies at $z \sim
0.35$ has however provided a new constraint by detecting a `baryon
acoustic peak'. Our best-fit E--deS models do possess a baryonic
feature at a similar physical scale as the best-fit $\Lambda$CDM
concordance model, but do not fit the new observations as well as the
latter. In particular the shape of the correlation function in the
range $\sim 10-100 h^{-1}$ Mpc cannot be reproduced properly without
violating the CMB angular power spectrum in the multipole range $l
\sim 100-1000$. Thus, the combination of the CMB fluctuations and the
shape of the correlation function up to $\sim 100 h^{-1}$Mpc, if
confirmed, does seem to require dark energy for a homogeneous
cosmological model based on (adiabatic) inflationary perturbations.}


\maketitle

\section{Introduction}

The detection by {\sl COBE} of large angular scale fluctuations in the
cosmic microwave background (CMB) opened a new era in modern cosmology
and established the inflationary model, in which the initial
conditions of the present Universe are set by fundamental physical
processes occurring in the very early Universe. The subsequent
detection of fluctuations on small angular scales and the considerable
improvement in the precision of their measurement in the last decade
has led to remarkable progress. The flatness of the universe was the
first key result to emerge from identification of the `acoustic peak'
in the angular power spectrum of the CMB (Lineweaver et al, 1997)
which allowed a measurement of the angular distance to the last
scattering surface.  Together with the observed low matter density of
the universe and the cosmic acceleration implied by the Hubble diagram
of Type Ia supernovae (Riess et al, 1998; Perlmutter et al. 1999),
this led to the `concordance model' of cosmology in which the dominant
component of the universe is a mysterious dark energy which behaves
essentially like a cosmological constant (see e.g. Peebles and Ratra
2003).

This model has been remarkably successful at fitting a large body of
cosmological data, in particular of large-scale structure
(LSS). However it should be emphasized that {\em direct} evidence for
dark energy, through the detection of the expected correlations
 between the CMB and LSS induced by the late integrated Sachs--Wolfe
effect, still has rather weak ($<3\sigma$) significance (see
e.g. Boughn \& Crittenden 2004, Padmanabhan et al. 2005). Dimming of
distant supernovae by grey dust (Aguirre 1999, Goobar et al. 2002,
Vishwakarma 2005) or evolution of their progenitors (Drell et
al. 2000, Wright 2002) could well mimic the effects of acceleration in
the SN~Ia Hubble diagram (see Figure 7 in Riess et al. 2004). Moreover
there are many `degeneracies' in the fitting of cosmological models to
CMB and LSS observations which allow rather different parameter
combinations to fit the same data. Given the complete lack of
theoretical understanding of dark energy, this motivated us to
reexamine (Blanchard et al. 2003, BDRS hereafter), whether
Einstein--de Sitter (E--deS) models could still be compatible with the
extant CMB and LSS data. This required assumptions at odds with
current beliefs: we adopted a low value of the Hubble constant (46
km/s/Mpc) and, motivated by theoretical ideas about inflation (Adams,
Ross \& Sarkar 1997, Martin \& Brandenberger 2003),\footnote{The
`glitches' appearing in WMAP and Archeops data may well be the
signature of such new physics (Ringeval \& Martin 2004, Hunt \& Sarkar
2004).} we relaxed the assumption that the spectrum of the primordial
density perturbation is a simple power-law. The fact that data from
WMAP as well as other CMB experiments can be reproduced with E--deS
models was a direct demonstration that the CMB data by themselves do
{\em not} require the presence of dark energy.

Moreover we showed that the power spectrum of large-scale structure as
measured in the {\sl 2dF} galaxy redshift survey, and inferred from
observations of the Lyman-$\alpha$ forest, can be adequately
reproduced if there is a small component (12\%) of unclustered matter
e.g. in the form of 0.8 eV mass neutrinos or a pressureless scalar
field. This lowers $\sigma_8$, the amplitude of matter fluctuations on
the scale $8h^{-1}$ Mpc, to 0.64 and 0.5 respectively, thus allowing
agreement with the value{\bf s} inferred from clusters and from weak lensing
for a critical matter dominated universe. In both models, the baryon
density is $\Omega_{\rm b}h^2 \simeq 0.02${\bf ,} in agreement with
the value inferred from primordial nucleosynthesis  (see Fields \&
Sarkar 2004 and Charbonnel \& Primas 2005 for recent discussions). 
Thus it appeared that an E--deS universe could indeed be
made consistent with all data, and possibly favored by the evolution
of the number density of distant clusters as inferred from {\sl XMM}
observations (Vauclair et al 2003) as well as the baryon fraction in
clusters (Sadat et al. 2005).

It had been emphasized by BDRS that further observations of
large-scale structure would provide a key test: {\it ``The most stable
difference between our E-deS models and the $\Lambda$CDM concordance
model is in fact the matter power spectrum shape in the range $k \sim
(0.01-0.03)~h$/Mpc, which galaxy surveys may be able to investigate,
provided the possible biasing is reliably understood on these
scales.''}  The recent results in this context from {\sl 2dFGRS} (Cole
et al. 2005), as well as {\sl SDSS} (Eisenstein et al. 2005) have
prompted us to return to this question.

\section{Galaxy Correlation functions in  Einstein--de Sitter models}

\begin{figure}
\centering 
\includegraphics[width=8cm]{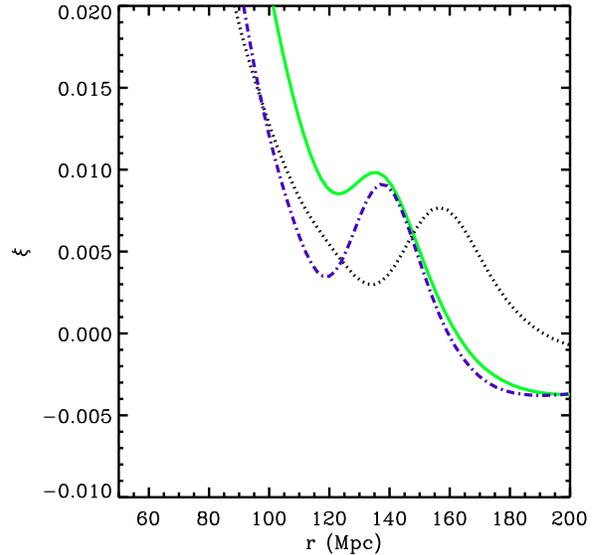} 
\caption{The correlation function in physical space for the best-fit
 power-law $\Lambda$CDM model (dotted, black line) from Spergel et al.
 (2003), and for the best-fit Einstein-de Sitter models $\Omega_\Lambda
 = 0$) from Blanchard et al. (2003) having a neutrino component
 $\Omega_\nu = 0.12$ (solid, green line) and a pressureless ($w =
 0$) quintessence component $\Omega_Q=0.12$ (dot-dashed, blue line). An
 acoustic peak is seen for all models, having a similar amplitude and
 on a similar {\em physical} scale, as can be anticipated from the
 physical origin of this peak.}
\label{ksiR}
\end{figure}

The most remarkable result that has emerged is the detection of the so
called  `baryon acoustic peak' in the angular correlation function of
{\sl SDSS} luminous red galaxies (LRGs), as well as in the {\sl
2dFGRS} power spectrum. The presence of such a feature is a robust
prediction (Peebles \& Yu 1970) and arises from the summation of
several wiggles in the power spectrum (see e.g. Blake \& Glazebrook
2003, Matsubara 2004, Huetsi 2005). The first question to examine is
whether the position of the peak discriminates between the concordance
$\Lambda$CDM model and the alternative E--deS models. We have
therefore computed the (dark matter) correlation functions of the two
best-fit E--deS models discussed by BDRS; these were obtained as
direct Fourier transforms of the corresponding power spectra without
taking into account corrections due to non-linear effects, redshift
space distortions, bias on large scales etc, which are in any case
expected to have little effect on the properties of the acoustic peak
(White 2005). The correlation functions were normalized to the
amplitude of the {\sl SDSS} LRG correlation function on the scale $10
h^{-1}$ Mpc and are shown in Figure~\ref{ksiR}. As one can see the
E--deS models exhibit a peak of similar amplitude and location in {\em
physical space} as the $\Lambda$CDM model, the reason for this being
that the cold dark matter density $\Omega_{\rm m} h^2$ and the baryon
density $\Omega_{\rm b} h^2$ have similar values in all these models.
 Note that the characteristic length scales as $(\Omega_{\rm b}
h^2)^{-0.252}(\Omega_{\rm m} h^2)^{-0.0853}$ (Hu 2005) making a $\sim
10 \%$ difference between our fiducial models.

\begin{figure}
\centering 
\includegraphics[width=8cm]{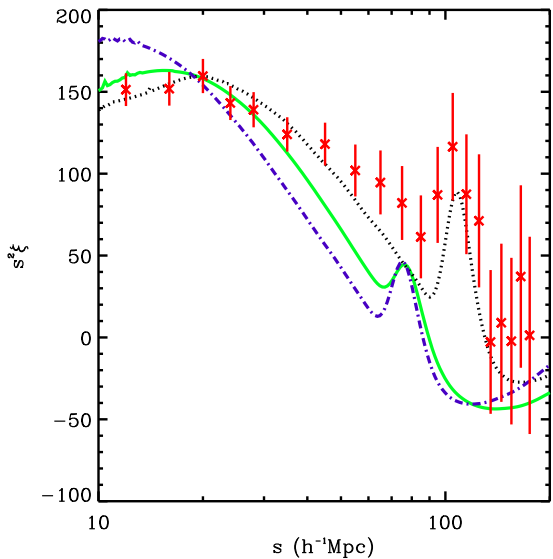} 
\caption{The correlation function in observed (redshift) space for the
same models as in Figure \ref{ksiR}, but with amplitude adjusted to
the best fitting value. For Einstein--de Sitter models, the spatial
scales are shifted for the geometrical factor (according to eq. [2]
of Eisenstein et al. 2005) relative to a $\Lambda$ dominated
cosmology. The position of the acoustic peak (in $h^{-1}$Mpc) for
E--deS models does not match the observations, unlike the $\Lambda$CDM
model, because a lower value of the Hubble constant is needed to fit
the CMB data. In addition, E--deS models exhibit a lack of power on
large scales (30--100 $h^{-1}$Mpc) when normalized optimally to the
observations.}
\label{S2ksi}
\end{figure}

However, the position of this peak in {\em redshift space} differs
because of the lower value of the Hubble constant needed to reproduce
the CMB power spectrum for the E--deS models. Because of this the peak
is shifted to smaller scales in redshift space as seen in
Figure~\ref{S2ksi}.  The abscissae of the E--deS models were rescaled by
a numerical factor of 1.22, taking into account the dilation at $z
\sim 0.35$ in distances according to eq. [2] of Eisenstein et
al. 2005).  Although the significance of this peak is not very high,
and needs to be confirmed in subsequent surveys, the  Einstein--de
Sitter models are significantly worse fits (having $\chi^2 \sim 95$
and $\chi^2 \sim 155$ for models in which the bias is optimized, { non
diagonal terms in the correlation matrix being neglected}) than the
concordance model ($\chi^2 \sim 20$) { for 19 d.o.f. Our concordance
model was previously determined as an optimal model for the CMB and is
therefore slightly worse than the optimal model of Eisenstein et
al. (2005) adjusted directly on the LRG data.}

For the $\Lambda$CDM model, the location of the peak { was 
predicted correctly which should be considered a great success. 
However, } its shape does not appear to be a particularly good fit to
the observations. This is also apparent from Figure~3 in Eisenstein et
al. (2005): models which reproduce well the amplitude of the peak have
an excessive amplitude on smaller scales ($20-50h^{-1}$Mpc). This
opens up the possibility that the origin of the feature is actually
different, perhaps imprinted in the primordial density
perturbation. Although this might appear unlikely, physical mechanisms
have been proposed which would generate such wiggles in the primordial
perturbation spectrum during inflation (Ringeval \& Martin 2004, Hunt
\& Sarkar 2004). It is not our intention here to argue in favor of
such a possibility,  but rather to examine
whether the shape of the correlation function itself can add new
information without any prejudice as to the nature of the primordial
density perturbation. As BDRS have shown previously, the CMB angular
power spectrum at $l \sim 1000$ is directly related to the amplitude
of matter fluctuations on the scale $\sim 10 h^{-1}$Mpc, therefore the
high precision CMB measurements impose a tight constraint on
$\sigma_8$ in a relative model independent way. Because of the smaller
amplitude of the power spectrum on large (redshift) scales, we expect
the correlation function in redshift space to reflect this lack of
power, as is seen in Figure~\ref{S2ksi} for the E--deS models we have
previously investigated. Because the amplitude of the CMB anisotropy
is well measured in the angular range $l \sim 200-1000$ (see
e.g. Jones et al. 2005), it follows that the amplitude of the matter
correlation function in the corresponding spatial range $\sim 30-100
h^{-1}$Mpc cannot be increased without exceeding the observed
amplitude of CMB anisotropy on those scales.

We have tried several modifications of the primordial power spectrum
but without success in matching the concordance model's ability to
reproduce simultaneously the data on CMB anisotropies and the LRG
correlation function. There are of course many unknowns apart from the
shape of the primordial density perturbation, for instance the
possible presence of inflationary tensor modes and of isocurvature
modes, the presence of topological defects, the nature of the dark
matter etc. It may not therefore be possible to rigorously rule out
models in which the CMB anisotropies and the LRG correlation function
can be reproduced without any dark energy. However it does appear that
the above argument is new evidence in favor of the presence of dark
energy, in the standard framework of general relativity and
homogeneous cosmological models.

\section{Conclusions}

Our main conclusion is that the new information contained in the {\sl
SDSS} LRG correlation function, in conjunction with observations of
CMB anisotropy, in principle allows discrimination between
cosmological models with and without dark energy. Although we have
shown earlier that Einstein--de Sitter models can indeed fit most
cosmological data (with the exception of the SNIa Hubble diagram and
the Hubble Key Project value of the Hubble constant), it now appears
possible to exclude them by more precise measurements of the
correlation function of galaxies on large scales, provided that
biasing is well understood. The concordance model does not have any
basis yet in fundamental physics and should therefore be regarded as a
convenient parameterization of the data in the context of the standard
FRW cosmology, rather than as a `Standard Model'. In particular there
is no physical explanation of why the universe should be embarking an
a new inflationary period at this late stage in its
history. Nevertheless we acknowledge that the new observations of the
galaxy correlation function, jointly with small-angle anisotropies in
the CMB which probe the same scales of $\sim 10-100 h^{-1}$ Mpc,
provide a remarkable geometric test of the concordance model which it
passes successfully.
   
\begin{acknowledgements}
We wish to thank Raul Angelo, Shaun Cole, Carlos Frenk and Paul Hunt
for helpful discussions. { We acknowledge useful comments from the referee.
MD acknowledge financial support from the French space agency CNES.}
\end{acknowledgements}

\end{document}